# A Multi-level Compiler Backend for Accelerated Micro-kernels Targeting RISC-V ISA Extensions

Alexandre Lopoukhine
University of Cambridge
Cambridge, United Kingdom
sasha.lopoukhine@cl.cam.ac.uk

Federico Ficarelli
University of Bologna
Bologna, Italy
Cineca
Bologna, Italy
f.ficarelli@cineca.it

Christos Vasiladiotis
University of Edinburgh
Edinburgh, United Kingdom
c.vasiladiotis@ed.ac.uk

Anton Lydike
University of Edinburgh
Edinburgh, United Kingdom
anton.lydike@ed.ac.uk

Josse Van Delm
KU Leuven
Leuven, Belgium
josse.vandelm@kuleuven.be

Alban Dutilleul
ENS Rennes
Rennes, France
alban.dutilleul@ens-rennes.fr

Luca Benini
ETH Zurich
Zurich, Switzerland
University of Bologna
Bologna, Italy
luca.benini@unibo.it

Marian Verhelst
KU Leuven
Leuven, Belgium
marian.verhelst@kuleuven.be

Tobias Grosser
University of Cambridge
Cambridge, United Kingdom
tobias.grosser@cst.cam.ac.uk

## Abstract

High-performance micro-kernels must fully exploit today's diverse and specialized hardware to deliver peak performance to deep neural networks (DNNs). While higher-level optimizations for DNNs are offered by numerous compilers (e.g., MLIR, TVM, OpenXLA), performance-critical micro-kernels are left to specialized code generators or handwritten assembly. Even though widely-adopted compilers (e.g., LLVM, GCC) offer tuned backends, their CPU-focused input abstraction, unstructured intermediate representation (IR), and general-purpose best-effort design inhibit tailored code generation for innovative hardware. We think it is time to widen the classical *hourglass* backend and embrace progressive lowering across a diverse set of structured abstractions to bring domain-specific code generation to compiler backends. We demonstrate this concept by implementing a custom backend for a RISC-V-based accelerator with hardware loops and streaming registers, leveraging knowledge about the hardware at levels of abstraction that match its custom instruction set architecture (ISA). We use incremental register allocation over structured IRs, while dropping classical spilling heuristics, and show up to 90% floating-point unit (FPU) utilization across key DNN kernels. By breaking the backend hourglass model, we reopen the path from domain-specific abstractions to specialized hardware.

## 1 Introduction

Modern general-purpose compiler frameworks use a mid-level, target-agnostic, RISC-like intermediate representation (IR) as input to a generic and well-optimized backend. As a result, C/C++, Fortran, Swift, Rust, and many more languages that target LLVM IR [52] all benefit from shared mid-level optimizations and mature CPU backends. While modern domain-specific language (DSL) compilers based on MLIR [53], effectively optimize deep neural networks (DNNs) in machine learning (ML) (and other domains) by progressively lowering across domain-specific abstractions, they commonly target LLVM as a backend. LLVM can indeed target CPUs reasonably well, but its load-store/arithmetic/branch-focused IR, with good support for single instruction multiple data (SIMD) operations, fails to effectively model the domain-specific structure of modern hardware. Intel uses a handwritten code generator [41] when targeting optimized SIMD for their CPUs, offering more control than the LLVM backends. As domain- and hardware-specific optimizations are hard to express in traditional backends, many domain-specific compilers, languages and libraries [5, 8, 27, 32, 71] commonly sidestep the *hourglass* backend (Figure 1–left).

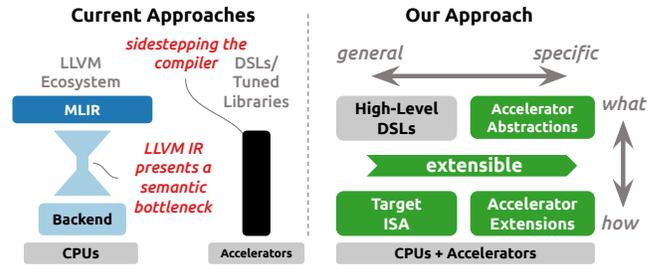

**Figure 1.** Traditional backends offer a narrow interface to abstraction-rich frameworks like MLIR, often triggering the development of expert-tuned libraries to exploit custom accelerator features (left). Our multi-level backend combines target-specific abstractions with progressive lowering offering a wide backend for accelerator code generation (right).

We widen this hourglass by proposing a multi-level backend which accepts, preserves, and exploits domain-specific information to target innovative hardware. Instead of a single catchall representation, our extensible design (Figure 1–right) exposes a family of static single assignment (SSA) IRs. We pair base IRs modelling the core instruction set architecture (ISA) (e.g., RISC-V) with structured IRs for domain-specific accelerator extensions. We support structured control flow and non-standard register usage through a multi-level register allocator that operates across IR abstractions and demonstrate that spilling common in best-effort register allocation is not needed for peak performance. By lowering from a high-level DSL, we show that a wide backend allows for direct lowering of domain-specific concepts to corresponding hardware features simplifying code generation compared to traditional backends. We implement our idea as a backend for Snitch [80], a scalable in-order RISC-V core with custom extensions for streaming registers and hardware loops, and generate high-performance DNN kernels for Snitch.

Our contributions are:

- A low-level representation of the RISC-V ISA and Snitch accelerator extensions in a collection of SSA-based IRs (Sections 3.1 and 3.2).
- A multi-level and spill-free register allocator for structured backend IRs (Section 3.3).
- A progressive lowering from a high-level DSL down to the aforementioned RISC-V SSA-based IRs (Section 3.4).
- An evaluation demonstrating peak performance for handwritten micro-kernels (up to 95% floating-point unit (FPU) utilization), and close-to-matching performance (up to 90% FPU utilization) for a high-level linear algebra DSL, when targeted via a micro-kernel compiler (Section 4).

## 2 Background

We describe foundations underpinning SSA-based compilers, advances in modeling high-level domain-specific concepts in SSA-based IRs, and present Snitch [80], a state-of-the-art RISC-V-based accelerator with custom ISA extensions. We leverage these concepts in our prototype backend, showing a novel and effective abstraction lowering for Snitch.

### 2.1 Static Single Assignment with Regions

Static single assignment (SSA) intermediate representations (IRs) are widely-used across modern research and industrial compilers (e.g., LLVM [52], GCC [3], Cranelift [1], xDSL [14]), thanks to the broadly-accepted benefits that explicit data flow information offers [22, 28]. SSA simplifies and improves analyses by forgoing the need to prove facts in every code location [22]. *Values* in SSA form are associated with a *unique* name, and each use of a value refers to a *unique* definition. We use SSA IR as implemented in MLIR [53].

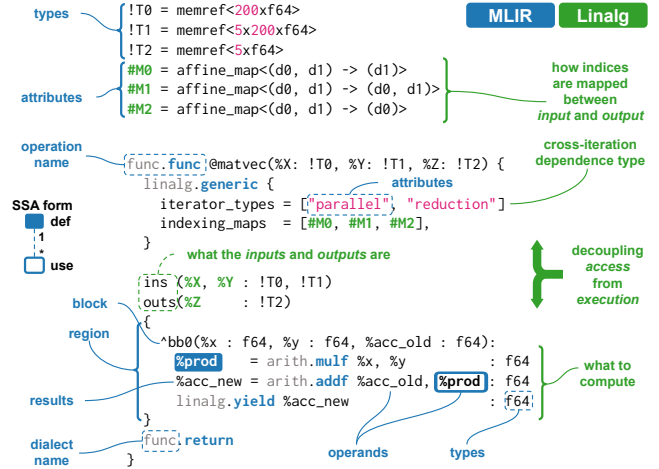

**Figure 2.** Organizing program abstractions as SSA-based IRs enables a modular approach for compiler construction. The above vector-matrix product in MLIR makes the use-def relationships explicit and obviates the need for intricate analyses by capturing information at the right abstraction level (e.g., directly expressing iteration types in linalg.**generic**).

*Operations* outline computation alongside their SSA values (i.e., results and operands) (Figure 2). A *dialect* forms a namespace for a set of related operations. Operations are prefixed with their dialect name (e.g., arith.**addf**) and may contain *attributes*, a key-value map of compile-time constants.

Operations organized in *blocks* correspond to straight-line code (i.e., basic blocks [28]). Blocks can represent bodies of functions or **for** loops, and may have values as arguments. A *region* is a list of blocks associated with an operation. Complex control flow between regions and blocks is defined by the semantics of their parent operation. For instance, scf.**for** embodies a typical **for** loop, with an induction variable incrementing within an integer range (Figure 2). Combining regions as a first-class IR element with SSA allows the direct encoding of nested hierarchical structures in the IR, without any restrictions in the combination of dialects used to express a program.

We use existing dialects that model common programming abstractions such as functions (func), structured control flow (scf) and memory reference manipulation (memref). For example, vector-matrix products can be implemented as a combination of these dialects (Figure 2). Functions passing arguments by reference are represented as func.**func** operations with memref arguments. We build on top of these abstractions, as well as existing machine learning DSLs to generate performant linear algebra micro-kernels (Section 4).

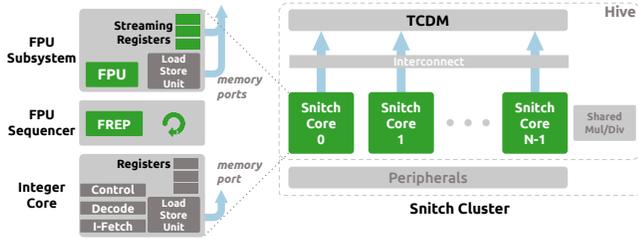

**Figure 3.** Simplified high-level overview of the Snitch microarchitecture used in our evaluation (Section 4) based on [80]. FPU utilization can be maximized using hardware loops (FREP) to remove explicit loop control flow and SSRs to eliminate explicit FP load/stores for affine access patterns.

### 2.2 Machine Learning Abstractions in SSA

SSA is well-suited in expressing programs with IRs at different abstraction levels for optimizations in compilation flows in ML [16, 55, 59] and other domains [19, 38]. The linalg dialect, an SSA-based IR, is a common lowering destination for high-level, ML-oriented IRs (e.g., onnx, pytorch) [12, 45].

This dialect concisely captures high-level linear algebra computations using a versatile operation, linalg.**generic**, encoding the following properties [58]: i) explicit iterator types, ii) affine mappings between iteration space and operand data, iii) an iteration space completely defined by input/output operands, and iv) a lambda specifying the computation (Figure 2). These properties are hard, or impossible, to reconstruct from low-level encodings [7, 17, 73].

### 2.3 RISC-V: An Extensible ISA

As technological challenges drive hardware designs towards vertical specialization, the RISC-V modular, extensible, and royalty-free instruction set architecture (ISA) is growing in popularity for domain-specific accelerators [42]. The RISC-V ISA defines a simple load-store reduced instruction set computer (RISC) architecture [75]. The ISA is organized in small groups of logically related instructions, simplifying their hardware implementation and enabling composition. Custom extensions are increasingly adopted to ship specialized designs [29, 60, 78] ranging, for example, from multicore cloud CPUs with hardware barriers, cache control and custom vector instructions [4, 15] to energy efficient architectures with multi-precision packed SIMD instructions [34, 63].

Increasing hardware specialization creates challenges in efficient code generation for traditional compiler backends. This is due to the widening gap between the high-level, application-driven semantics of such extensions that are hard to represent with common IR and backend data structures [57].

### 2.4 The Snitch Architecture

Snitch [80] is an open-source, state-of-the-art RISC-V core design by ETH Zurich. It applies novel architectural solutions to address the compute scalability challenge in terms of energy efficiency, achieving more than double the performance per Watt of other leading commercial accelerators. It has been successfully used as the fundamental processing element (PE) of large, multicore accelerators, like Occamy [61], Manticore [79] and heterogeneous tiled designs [2].

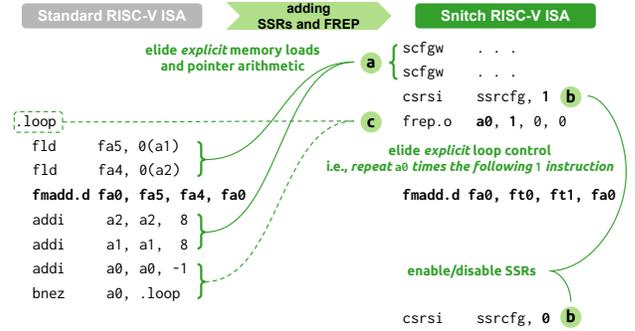

**Figure 4.** Introducing the Snitch ISA extensions (right) to the implementation of a vector product kernel in the standard RISC-V ISA (left) allows the elision of explicit loop control operations (pointer arithmetic and branching) and memory operations, maximizing FPU utilization by decoupling FP instruction processing from other instructions within the Snitch in-order core.

Snitch comprises a lean, in-order integer core serving a large floating-point unit (FPU), capable of multi-precision and (non-standard) packed SIMD instructions [34, 67]. The integer core connects to the floating-point (FP) subsystem through a sequencer unit, which drives FPU operations independently (Figure 3). Snitch also uses a fast, energy-efficient and high-throughput tightly-coupled data memory (TCDM) (128 KiB), acting as a software-managed L1 cache. Its *key* design idea is to maximize area and energy efficiency, by utilizing the FPU core for the majority of the computation. Two custom features enable optimal FPU utilization:

- **Stream semantic registers (SSRs)** implicitly handle FP loads/stores when adhering to affine memory accesses [65].
- **Floating-point repetition (FREP)** repeatedly executes an FP instruction sequence, removing the need for loop control flow (conditionals, jumps, induction variables).

SSRs are semantically similar to vector operations, removing the need for loads/stores from/to TCDM (Figure 4: a b ). Using FREP allows the FPU to execute instructions without waiting for the integer core to provide them, effectively making the architecture pseudo-dual issue (Figure 4: c ). SSRs and FREP can efficiently handle dense linear algebra operations without over-specialization.

The Snitch packed SIMD instructions [33] predate and, thus, differ from those ratified in the standard RISC-V P extension specification [10]. Snitch registers are distributed over eight lanes, allowing simultaneous operation on eight 8-bit, four 16-bit or two 32-bit values per 64-bit-wide register.

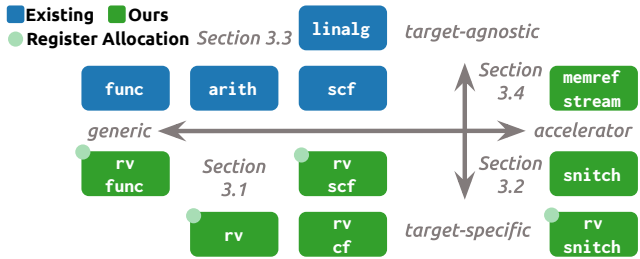

**Figure 5.** Our multi-level compiler backend utilizes a host of MLIR dialects to generate efficient code, tailored to the RISC-V Snitch accelerator. The high-level information from the `linalg` dialect is progressively lowered to the custom Snitch ISA extensions. Our modular approach enables the partitioning of challenging tasks, such as register allocation, at a suitable level of abstraction.

Implementing Snitch extensions in traditional compilers has proven challenging, as its capabilities introduce implicit memory accesses and iterations, requiring reconstruction of code guarantees from backend analyses [6, 80].

## 3 A Multi-Level Compiler Backend

We present the architecture of a multi-level compiler backend in four steps (Figure 5). We introduce MLIR dialects that model RISC-V assembly and higher-level target-specific concepts, enabling high-level reasoning in the backend (Section 3.1). We then discuss how Snitch ISA extensions can be represented by additional dialects, building naturally on top of the base RISC-V infrastructure (Section 3.2). We show how maintaining structured control flow in our backend allows for more direct and predictable register allocation (Section 3.3). Finally, we demonstrate how a progressive lowering from domain-specific operations to multi-level compiler backend abstractions enable generation of code that effectively harnesses the accelerator capabilities (Section 3.4).

### 3.1 SSA-based IRs for the RISC-V ISA

Our RISC-V backend represents target-specific concepts at multiple levels of abstraction, leveraging the MLIR infrastructure throughout the compilation flow. We model assembly instructions with the lower level `rv` and `rv_cf` dialects and encode semantics (e.g., structured control flow in `rv_scf` and call conventions `rv_func`) with the higher ones (Figure 5). In contrast to monolithic backends, our infrastructure is split into components that are easy to reason about and extend.

We define the `rv` dialect using MLIR's extensible type system, denoting assembly instructions as operations where source and destination registers correspond, respectively, to operands and results (Figure 6: 1 ). Some operations, such as `rv.get_register`, are not printed in the assembly; these exist to create SSA values in the IR, bridging SSA semantics and our representation of registers in types (Figure 6: 2 ). Operations in the `rv_cf` dialect model unstructured control flow via jump instructions to other basic blocks in the IR. Assembly is printed using an interface-based design, where the IR is walked in-order, and printed according to implementation of each operation. Our low-level program representation allows for compiler reasoning (e.g., control flow) and transformations while keeping a close correspondence to the target assembly.

Higher level RISC-V dialects let us preserve more semantic information that is useful for target-specific optimizations. For example, the `rv_func.func` operation (Figure 6: 3 ) encodes the application binary interface (ABI) constraint of requiring function arguments and results to be passed in *A* registers. Similarly, the `rv_scf.for` operation represents a for loop in a structured way, easing optimizations and live range construction during register allocation. These dialects are designed to mirror the existing `func` and `scf`, making lowering from higher abstractions straightforward.

### 3.2 Representing RISC-V ISA Extensions

In contrast to monolithic compiler backends, our modular approach eases the addition of new capabilities. We augment our backend structure for ISA extensions by following a similar multi-level approach, explicitly encoding accelerator semantics and application programming interfaces (APIs) in the IR. Our high-level operations let us easily optimize and reason about accelerator usage.

We model the Snitch packed SIMD operations, streaming configuration, and floating-point repetition (FREP) loops (Section 2.4) in `rv_snitch`. stream semantic registers (SSRs) add memory effects to previously pure arithmetic operations, which we model with operations that explicitly interact with streaming registers (Figure 6: a ), a representation that lets us decouple generic analyses on data access from those on downstream processing. We represent packed SIMD instructions similarly to the standard FP instructions as these operate on scalar FP registers. We model FREP with a region body and iteration count operand, along with a mechanism to accumulate results (Figure 6: b ), adding the constraint that only instructions on FP registers and stream operations are allowed in the loop body. We define the `snitch_stream.streaming_region` operation to encapsulate the streaming configuration and the region where streaming is enabled (Figure 6: c ). These operations provide convenient targets for higher-level compiler passes.

Our representation of stream configurations as compile-time constants allows us to easily perform some key performance optimizations (Figure 6: d ). A single construct for upper bounds and strides allows us to detect and remove contiguous accesses, reducing the number of generated assembly operations for accelerator configuration. Similarly, a stride of 0 in the last dimension represents a repeated memory access to the same location, for which the Snitch ISA extensions provide a dedicated optimization, reducing the pressure on the memory interconnect in the hardware. Declarative, high-level representations of accelerator capabilities allow the compiler to use simple peephole rewrites for custom optimizations.

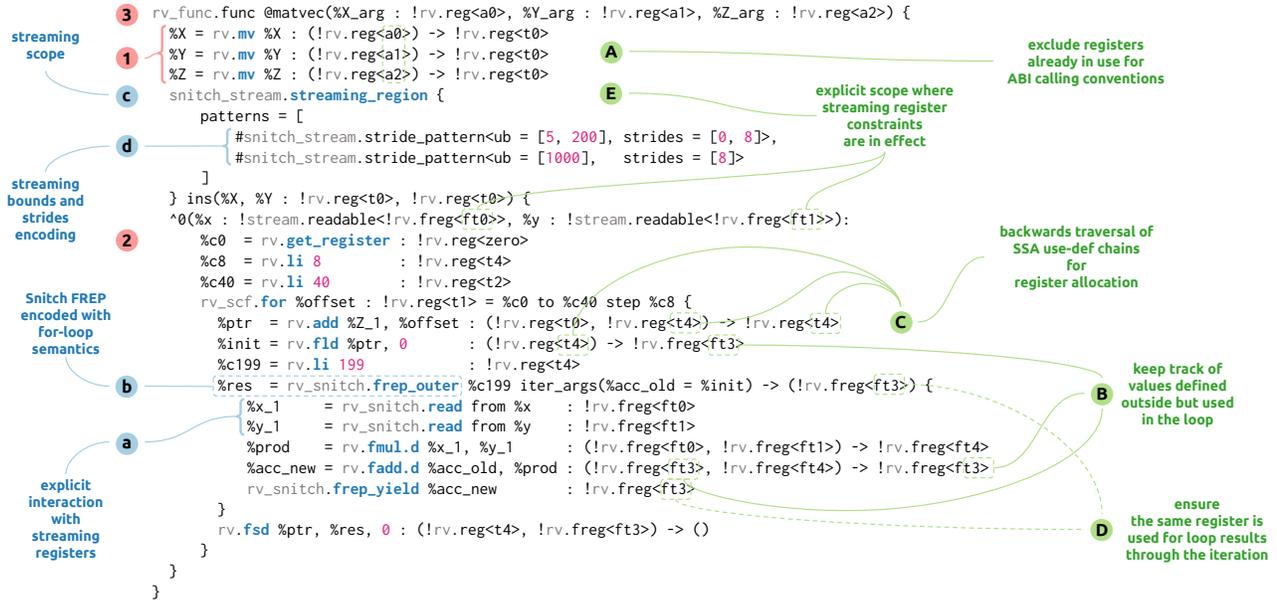

**Figure 6.** Our multi-level backend uses a mix of SSA-based IRs to represent different levels of abstraction around the RISC-V ISA for a matrix-vector calculation. The SSA formulation of the ISA empowers the compiler to employ well-understood analyses and transformations and, when combined with regions, to encode further information control flow information (e.g., for loops) while staying close to the semantics of the ISA.

Our modular approach makes it easy to extend backends to target accelerators with custom ISA extensions, by leveraging shared abstractions and infrastructure.

### 3.3 Register Allocation in SSA with Regions

We present a spill-free register allocation approach that leverages information such as structured control flow, explicitly encoded in our high-level RISC-V and Snitch IRs, to reason about liveness of values.

We allocate registers in three linear passes. The first pass excludes all registers already used in the IR from the 15 integer (a and t) and 20 FP registers (fa and ft) that are specified as caller-saved in the RISC-V ABI (Figure 6: A ). By excluding used registers, we can process partially-allocated code in a generic way, even if the exclusion of the registers is overly defensive, as a precise calculation of the live ranges of pre-allocated values would add complexity to our design. The second pass keeps track of variables defined outside the region in which they are referenced, for later use when processing loops (Figure 6: B ). The final pass walks the IR backwards and allocates values in-place, assigning registers on first use, and freeing them on definition (Figure 6: C ). SSA form guarantees that a linear walk respects the order of use-def relations; this property extends to MLIR's SSA with regions, letting us allocate whole function bodies in a single walk of their implementation.

Loop iteration results are represented as operands before iteration, and block arguments during iteration, which we allocate first to make sure their registers match (Figure 6: D ). We then allocate the variables tracked in the second pass, as mentioned above; the live ranges of values used in a loop body, but defined outside, are extended beyond its scope, due to the possibility of executing the loop multiple times. With these values allocated, we recursively process the loop body with a backwards walk. Recursion lets us allocate arbitrarily-nested loops, and separates the logic of loop and single-block processing. Our loop processing approach is a simple extension to the single-block steps outlined above.

The Snitch ISA extensions impose additional constraints on register allocation. Snitch reserves the use of some registers during streaming, which we express in a declarative way in the relevant operation definition (Section 3.2) (Figure 6: E ). Similarly, the FREP operation declares its loop structure explicitly, and is processed in the same way as the rv_sc.for loop. The generic design of the allocator reduces the effort required to implement micro-kernel compiler backends.

Register allocation in generic compiler backends is typically performed on code at a low level of abstraction, where analysis of basic block graphs is needed to reconstruct liveness and control flow [28, 77]. The handling of unstructured control flow is out of scope when the input to the compiler is guaranteed to be in a linear algebra DSL. Similarly, spilling,

```
!T0 = memref<200xf64>
!T1 = memref<5x200xf64>
!T2 = memref<5xf64>
!S = !memref_stream.readable<f64>
#M0 = affine_map<(d0, d1, d2) -> (d1)>
#M1 = affine_map<(d0, d1, d2) -> (d0 * 5 + d2, d1)>
#M2 = affine_map<(d0, d1) -> (d0 * 5 + d1)>
#SP0 = #memref_stream.stride_pattern<ub = [1, 200, 5], index_map = #M0>
#SP1 = #memref_stream.stride_pattern<ub = [1, 200, 5], index_map = #M1>
memref_stream.streaming_region {patterns=[#SP0,#SP1]} ins(%X,%Y:!T0,!T1) {
^0(%0 : !S, %1 : !S):
  memref_stream.generic {
    bounds = [1, 200, 5],
    indexing_maps = [#M0, #M1, #M2],
    iterator_types = ["parallel", "reduction", "interleaved"]
  } ins(%0, %1 : !S, !S) outs(%Z : memref<5xf64>) {
  ^1(%x0 : f64, %x1 : f64, %x2 : f64, %x3 : f64, %x4 : f64,
     %y0 : f64, %y1 : f64, %y2 : f64, %y3 : f64, %y4 : f64,
     %a0 : f64, %a1 : f64, %a2 : f64, %a3 : f64, %a4 : f64):
      %b0 = arith.mulf %x0, %y0 : f64
      %b1 = arith.mulf %x1, %y1 : f64
      %b2 = arith.mulf %x2, %y2 : f64
      %b3 = arith.mulf %x3, %y3 : f64
      %b4 = arith.mulf %x4, %y4 : f64
      %c0 = arith.addf %a0, %b0 : f64
      %c1 = arith.addf %a1, %b1 : f64
      %c2 = arith.addf %a2, %b2 : f64
      %c3 = arith.addf %a3, %b3 : f64
      %c4 = arith.addf %a4, %b4 : f64
      memref_stream.yield %c0, %c1, %c2, %c3, %c4
                    : f64, f64, f64, f64, f64
  }
}
```

**Figure 7.** The `memref_stream` abstractions bridge the gap between high-level linear algebra abstractions and Snitch accelerator capabilities, allowing us to schedule computation before separating access from execution.

a feature required for general-purpose register allocation, has a negative performance impact, making it undesired for micro-kernel compilation. Our structured approach simplifies register allocation by preserving the relevant information present in the source into the backend.

### 3.4 Targeting High-Level Backend Abstractions

Our compiler is designed with the guiding principle of leveraging our knowledge of the target hardware in combination with information provided in the source as early as possible. We use the decoupled structure of memory accesses and computation in `linalg.generic` operations to target Snitch's N-dimensional access patterns and FREP loops. We design a schedule that is able to effectively target available hardware resources by rewriting high-level representations of micro-kernels in place, before lowering to control flow constructs. Our approach pushes a representation of accelerator hardware capabilities above the abstraction level that they are usually limited to in traditional compilers.

The `memref_stream` dialect bridges `linalg` abstractions and `snitch_stream` operations (Figure 7). The `memref_stream.generic` operation is based on its `linalg` counterpart, except an explicit encoding of the iteration bounds, in contrast to `linalg`'s approach of inferring bounds from input shapes (Section 2.2). This decoupling lets us model functions on values without shapes, such as streams. The `memref_stream.streaming_region` is the higher-level counterpart to the `snitch_stream` operation, operating on abstract arithmetic values instead of RISC-V registers (Section 3.2). As the streaming configuration fixes the order of data accesses, we must decide on a schedule before separating data access from execution when lowering.

Snitch's in-order core and software-managed L1 cache make its performance predictable, allowing us to build a simple scheduling pipeline, obviating the need for expensive schedule space exploration. To avoid accumulating intermediate results in memory, we exclude the reduction indices from the iteration space specifications of the results, guiding our lowering to loops to use local values for accumulation. Read-after-write (RAW) conflicts are averted by applying unroll-and-jam, which interleaves multiple iterations in the innermost loops, trading off increased code size and register pressure for performance (Figure 7). We automatically select the optimal unroll factor based on the pipeline depth. For Snitch, the FPU has three stages for all operations, so stalls are minimized when the unroll factor is at least four. The iteration space is fixed by these transformations, allowing us to separate the stream setup from the computation, which can be lowered to `scf` loops, with the inner operations operating on streams instead of memory. The lowering is structured as small, self-contained passes, making it easier to introspect, develop and maintain, and letting us reach close to optimal performance (Section 4.4).

Instead of discarding the high-level information and constraints of `linalg.generic` operations, and having to reconstruct them from loop representations in the backend, we lower them directly to custom operations representing a Snitch streaming bodies. A decoupled representation of access and execute in a single operation lets us deterministically construct schedules adapted to Snitch's capabilities, reducing the effort required for high-performance code generation for our target hardware.

## 4 Evaluation

We evaluate whether our multi-level backend enables effective kernel compilation by posing three research questions:

- **RQ1**: Are our assembly-level RISC-V dialects expressive enough to represent neural network micro-kernels tuned for peak performance?
- **RQ2**: Is a spill-free multi-stage register allocation approach suitable for generating such micro-kernels?
- **RQ3**: When targeted from a high-level DSL, can our backend still consistently generate peak-performant code?

After giving an overview of our experimental setup (Section 4.1), our results show: micro-kernels expressed in our assembly-level dialects can be effectively compiled to the Snitch accelerator (**RQ1**, Section 4.2) reaching up to 95% of the peak performance; our multi-stage register allocation approach is consistently spill-free (**RQ2**, Section 4.3); and our backend can be effectively targeted from a higher-level DSL (**RQ3**, Section 4.4) obtaining 90% FPU utilization.

**Table 1.** We evaluate our multi-level compiler on representative DNN micro-kernels (grouped by computational and memory access traits) across various input shapes. FLOPs indicates the minimum cycles needed for each computation.

| Kernel | Characteristics | Input Shapes | FLOPs |
|---|---|---|---|
| Sum<br>Fill | element-wise<br>linear access<br>memory-bound<br>parallel | $NM$, $NM$ | $NM$ |
| ReLU | element-wise<br>non-linear access<br>parallel | $NM$ | $NM$ |
| Conv 3×3 | non-affine access<br>fixed-size reduction | $(N+2)(M+2)$ | $18NM$ |
| Max Pool 3×3<br>Sum Pool 3×3 | sparse access<br>fixed-size reduction | $(N+2)(M+2)$ | $9NM$ |
| MatMul<br>MatMulT | nested loops<br>reduction | $NK$, $KM$ | $2NMK$ |

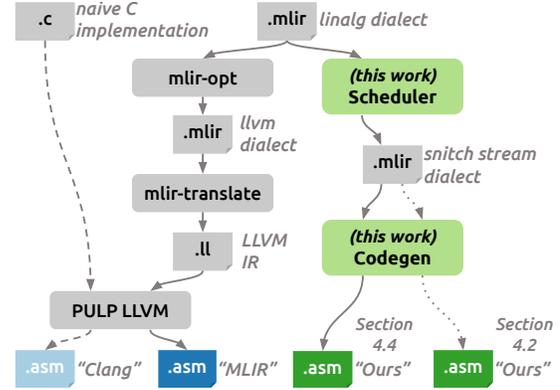

**Figure 8.** We compare our prototype compiler with flows using Clang and MLIR, and separately evaluate the expressivity of our MLIR backend (Section 4.1).

### 4.1 Experimental Setup

We compile representative neural network (NN) micro-kernels to Snitch, a state-of-the-art RISC-V-based accelerator, comparing against existing compilation flows. We now detail our kernels, inputs, simulation infrastructure, methodology, compilation flows and other software used.

***Kernels.*** We obtain micro-kernels from two DNNs NS-Net2 [23], a noise suppression model, and AlexNet [50], an image classification model. The two networks represent both a recent and a classical NN architecture. We manually implement the kernels (Table 1) used by these two networks, excluding Softmax and Sigmoid, as these rely on exponentiation and logarithm functions, the implementation of which is beyond the scope of this paper. Our selection of kernels includes element-wise computations, linear and non-linear memory accesses, reductions, and nested loops, covering a wide range of characteristics.

***Inputs.*** We select shape sizes to fit within the TCDM (Section 2.4) such that our performance measurements are not influenced by the rest of the memory hierarchy. Kernels with reductions (e.g., MatMul) comprise two linalg operations: i) zeroing out the output buffer (with linalg.**fill**), and ii) performing the computation. Across experiments, we evaluate a range of sizes to provide detailed information with respect to computational overheads.

***Simulation Target.*** As a state-of-the-art RISC-V-based accelerator we choose Snitch [80], which forms the core of one of the first chiplet-based research architectures [11]

(Section 2.4). We compile the open-source reference SystemVerilog implementation of Snitch with Verilator [13] to generate the register transfer level (RTL) *cycle-accurate* simulator. According to the Snitch micro-architecture, the FPU can execute one instruction per cycle peak or two floating-point operations (FLOPs) per cycle peak in case of the fused multiply-add (FMA) instruction), when operating on 64-bit values. For smaller types a corresponding number of vector operations can be executed. As an in-order, *bare-metal* platform with no runtime or operating system, all measurements on the Snitch platform are deterministic.

***Methodology.*** To obtain accurate performance data, we measure the cycle count, throughput, and FPU utilization of our kernels, leveraging Snitch's performance counters in combination with instruction trace postprocessing provided by our Verilator infrastructure. Our cycle count measurements for each kernel include the overhead of function calling, integer arithmetic instructions, explicit load/store instructions, and accelerator setup. We calculate the FLOPs for each kernel as the product of: i) loop iteration counts and, ii) the FLOPs performed in the loop body. We define *throughput* as the count of FLOPs per cycle during kernel execution. Together with the theoretical maximum throughput, the kernel FLOP count can be used to estimate the minimum number of cycles required to perform the computation. We count the FMA (fmadd) instruction as two FLOPs, and adjust our minimum cycle estimate for the kernels that use it. FPU utilization is expressed as the ratio of cycles spent in the FPU executing arithmetic instructions over the total execution latency. These three metrics assess the kernel execution speed and the *efficient* use of the available compute resources.

***Compilation Flows.*** General-purpose compilers such as LLVM can target the RISC-V ISA, generating code that can be executed on Snitch. We use two alternative compilation flows (Figure 8), both leveraging the LLVM RISC-V backend: i) a pipeline using existing MLIR passes, lowering the same inputs as our prototype compiler (tagged as "MLIR"), and ii) a naive C reimplementation of the same kernel (tagged as "Clang"). As no existing compilers target the Snitch ISA extensions, these flows are presented for additional discussion, and not a baseline for direct comparison.

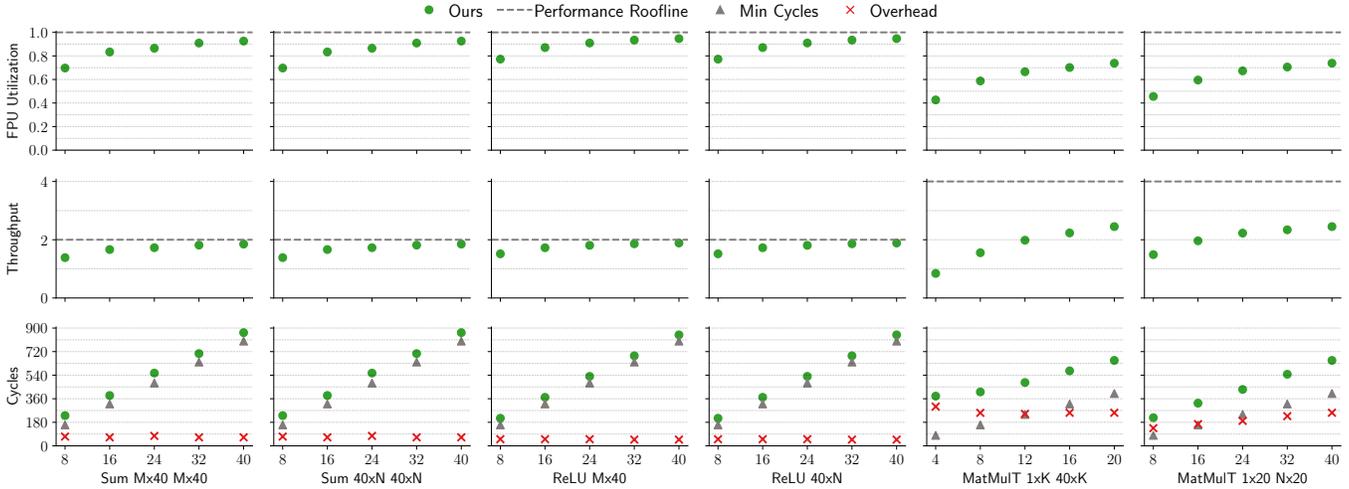

**Figure 9.** Our low-level representation is flexible enough to represent linear algebra operations commonly used in machine learning (ML) reaching high FPU utilization, reaching 95% peak FPU utilization and 94% of theoretical maximum throughput. Despite the high FPU utilization, the MatMulT kernel only reaches 2.45 throughput due to extra vector packing instructions.

***Software.*** We take advantage of MLIR's rich ecosystem of compiler components and tools. Our backend is implemented in the xDSL open-source compiler framework (v0.21.1) [14, 31], the *Pythonic* counterpart of MLIR, enabling native integration of core MLIR constructs (i.e., SSA, regions) within the language. Interoperability between xDSL and MLIR (v16.0.6) is achieved via the common text IR format. For C implementations, we use the LLVM toolchain provided by the Snitch architects [6], containing both the assembler and linker used in all our kernel implementations. Leveraging existing tools popular in industry eases the adoption of our work in research and production.

### 4.2 Low-Level Micro-kernel Representations

To assess the expressive power of our RISC-V dialects, we analyze selected micro-kernels (Figure 9), written in a combination of the RISC-V dialects (Section 3.1) and dialects encoding the Snitch ISA extensions (Section 3.2), expressed in a partially register-allocated form. Our measurements demonstrate effective FPU usage for all chosen kernels.

We express the Sum, ReLU, and MatMulT kernels on 32-bit FP data, using the `snitch_stream`, `rv_snitch` and structured `rv` dialects, and lower to assembly using our backend code generation passes. The Sum and ReLU kernels display similar performance behavior and attain 95% FPU utilization. These kernels are element-wise operations on one or two operands, have no reductions, and operate in linear manner, resulting in a minimal and constant overhead for the accelerator setup and a simpler control flow structure, reaching nearly 100% FPU utilization. The cycle count overhead remains constant independent of the sizes (bottom of Figure 9), suggesting that, subject to memory constraints of the Snitch cluster, the performance of the kernels will trend towards 100% of the theoretical roofline as the sizes increase. The MatMulT kernel reaches 74% FPU utilization, but only attains a throughput of 2.45 FLOPs/cycle. All three kernels match the performance of our best-effort optimized, handwritten assembly versions. We believe that our low-level representations of kernels are close to optimal, given the available extensions in the Snitch ISA.

Our low-level abstractions allowed us to express all chosen kernels, without posing any inherent obstacles that would require circumventing or supplementing them with external features, while reaching high performance on our target.

### 4.3 Spill-Free Register Allocation of Micro-kernels

The scarcity of on-chip accelerator resources complicates efforts to achieve high utilization, with registers being a prevalent constraint. Register spilling schemes are employed once register pressure exceeds the available physical register capacity, resulting in performance penalties from reaching into the target memory hierarchy. We examine the register use of our multi-level register allocator, which avoids spilling entirely, over a range of data and shape sizes, to determine its suitability in generating highly performant kernels (Table 2).

For the double-precision kernel variants, register pressure is manageable, with a surplus of unallocated registers. For instance, in the Fill kernel, we use two FP registers—one for the fill value and one for streaming the output—as well as two integer registers: one for the buffer pointer and another for the streaming configuration. The reserved registers bring the usage to three registers each for FP and integer types, since the fill value and the buffer are provided as function arguments to the kernel. The pooling, convolution, and matrix multiplication kernels are unrolled by a factor of four, and have multidimensional access patterns increasing the pressure for both integer and FP registers. In the double-precision case, the margin in available registers allows for more complex kernels while avoiding spilling.

**Table 2.** Our register allocator consistently manages the available 20 FP and 15 integer registers across all kernels, maintaining several spare. The higher complexity of the 32-bit kernels increases the register pressure.

| Kernel | Precision (bits) | Input Shape Sizes (#) | | | Allocated Registers (#) | |
| --- | --- | --- | --- | --- | --- | --- |
| | | N | M | K | FP | Integer |
| Fill | 64 | 4 | 4 | – | 3/20 | 3/15 |
| ReLU | 64 | 4 | 4 | – | 3/20 | 5/15 |
| Sum | 64 | 4 | 4 | – | 3/20 | 7/15 |
| Max Pool 3×3 | 64 | 4 | 4 | – | 7/20 | 6/15 |
| Sum Pool 3×3 | 64 | 4 | 4 | – | 7/20 | 6/15 |
| Conv 3×3 | 64 | 4 | 4 | – | 8/20 | 8/15 |
| MatMul | 64 | 4 | 16 | 8 | 8/20 | 8/15 |
| ReLU | 32 | 4 | 8 | – | 3/20 | 5/15 |
| Sum | 32 | 4 | 8 | – | 3/20 | 7/15 |
| MatMulT | 32 | 4 | 16 | 16 | 11/20 | 12/15 |

The single-precision variants have a higher range of register use. The 32-bit variant of MatMulT is the only kernel exhibiting high register pressure for both FP and integer types. This kernel computes the dot products of even and odd elements of rows from the input matrices using SIMD operations (Section 3.2), sums them up, and stores the result at the corresponding offset. Similarly to the other kernels, to achieve high FPU utilization, this kernel is unrolled by a factor of four. The 11 FP registers used are: i) 4 registers for the loop reduction values, ii) 4 registers for the result values, iii) 1 register for the initial zero value, and iv) 2 registers for streaming inputs. The integer registers used are either the three reserved function arguments, pointers and offsets used in the loop calculations, or other temporary values. Despite its complexity, this kernel requires no spilling.

In both the single and double precision kernels, FP registers are used predominantly for the streaming registers, constant data values and intermediate results, whereas the integer registers are used for pointers to buffers and loop bookkeeping. We reserve registers for the function arguments and results of a kernel, a choice that simplifies implementation of kernels but increases their register pressure (Section 3.3). While the decision to exclude registers used for function arguments and results does not present a challenge for kernels of similar characteristics to our chosen set, it might become a concern for those with more involved features such as deeper loop nests. This trend is supported by our analysis of kernels with a varying complexity of computation and memory access patterns (Table 2). Optimizations to mitigate this constraint, as well as other transformations to reduce register pressure, are planned for future work.

Some optimizations, like loop unrolling, increase code complexity and consequently register pressure (Section 4.4). Despite these factors, our measurements support our claim that spill-free register allocation is well-suited for linear algebra micro-kernels.

### 4.4 Prototype Micro-kernel Compiler

We evaluate the capacity of our multi-level compilation approach to generate target-optimized, high-efficiency kernels from high-level abstractions. We compare the performance of code generated by our compiler (Section 3.4) to code compiled with MLIR and a naive C implementation compiled with Clang (Section 4.1). As in Section 4.2, the goal is to minimize kernel execution time and maximize FPU utilization.

The MLIR and Clang compilation flows do not yield code that effectively exploits the capabilities of the hardware. In our measurements, they have similar performance, both peaking at approximately 42% utilization (Figure 10). The two flows share the LLVM RISC-V backend, which does not accommodate the characteristics of the in-order Snitch CPU or its ISA extensions. This results in suboptimal patterns in the generated assembly for both compilation flows, such as explicit loads/stores and RAW hazards, hence justifying the low performance. Max Pool benefits the most due to unrolling of some loops and rescheduling loads to minimize latencies by the LLVM backend, yet it remains below 50% FPU utilization. The *key* insight is that even when using high-level domain-specific MLIR optimizations, the LLVM IR and backend determine, and ultimately limit, the performance of the micro-kernels.

On the other hand, our approach results in high utilization even for the smaller sizes of our kernels, and reaches as high as 90% (Figure 10), regardless of the accelerator setup overhead. For the parallel kernels (Sum, Fill and ReLU), FPU utilization increases as input sizes grow, approaching 100%. For the reduction kernels (Conv, Max Pool and Sum Pool), when the input width varies, the FPU utilization increases, but at a slower rate, staying within the 70–80%. All our kernels are transformed to process four reductions at a time, leading to the outlier of ∼90% utilization when $N = 4$, as the (now single-iteration) outermost loop is unrolled, removing the loop overhead and reducing the number of dimensions in the accelerator setup. The rest of the kernels behave analogously to their parallel counterparts, increasing steadily in utilization as the width of the kernels increases.

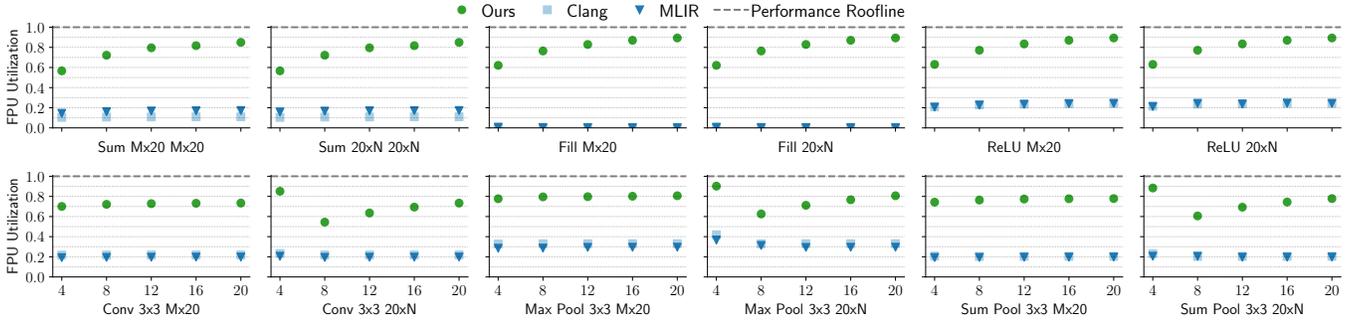

**Figure 10.** Selected micro-kernels compiled with our end-to-end prototype compiler reach up to 95% FPU utilization. In contrast, MLIR does not outperform a naive C implementation compiled with Clang on this platform.

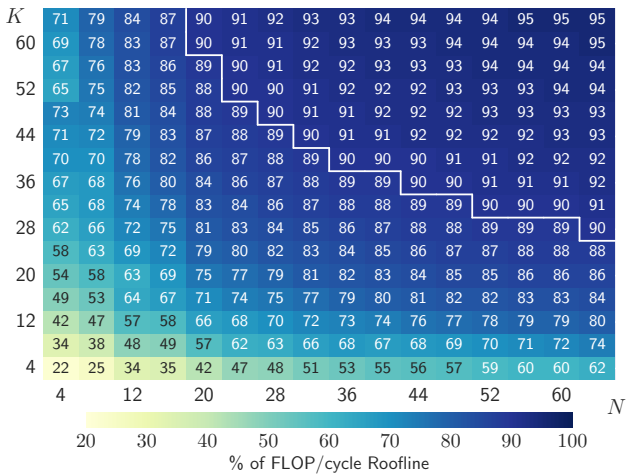

**Figure 11.** Sustained throughput of the 64-bit MatMul kernel ($C_{M \times N} = A_{M \times K} B_{K \times N}$ when $M = 1$). We achieve throughput of over 90% ($\geq 1.80$ FLOPs/cycle) of the theoretical peak (above the white border) as shape sizes increase, indicating that the computation offsets constant overheads.

We ran additional measurements of the MatMul kernel to examine its performance over a wider range of input shapes (Figure 11). For the smallest sizes of either the inner dimension or the columns of the second operand, the accelerator setup costs dominate the runtime, resulting in a relatively low throughput, never reaching above 80% of the theoretical peak. As with the other kernels, larger input shapes result in higher throughput, as the constant overheads of the setup and function calls are outweighed by the runtime of the computation. Similarly, throughput increases less rapidly as either input size gets larger. These trends should be taken into account by higher-level tools calling into our compiler when distributing larger workloads between Snitch cores.

To understand the impact of each high- and low-level optimization in our pipeline, we applied them incrementally on the MatMul kernel (Table 3). The kernel is composed of two linalg.**generic** operations; the first initializing the output to zero, and the second computing the matrix multiplication. Together, these operations constitute a non-accumulating kernel, which is the form used by most MLIR DNN frontends. The baseline pipeline represents direct lowering, with no schedule optimizations, targeting the standard RISC-V ISA. The resulting FPU occupancy is below 3%, an approximate 36× slowdown compared to the full pipeline.

The Streams optimization enables the use of Snitch SSRs, resulting in a 50% decrease in the cycle count. Meanwhile, the number of explicit loads and stores issued decreases by 66%, but the performance remains low. Adding Scalar Replacement leads to an additional performance improvement of over 4× by further reducing explicit loads and stores, since the kernel no longer accumulates the result directly into memory. Enabling FRep at this stage yields only a marginal improvement in the overall cycle count, as the run time is dominated by the calculation of the dot product in the innermost loop, and not the loop setup overheads. All remaining optimizations affect execution scheduling, the importance of target-aware schedules early on in the pipeline.

By fusing (Fuse Fill) the initial zeroing of the result matrix into the matrix multiplication operation, we reduce even more the number of instructions. In the case of a matrix multiplication with a large inner dimension, the computation dominates the time required to fill the result, so the performance gain is minimal. More importantly, after fusing the initialization with zeros we are able to ignore the previous contents of the matrix multiplication result buffer, eliminating the remaining loads and stores. At this stage, the limiting factor is the cross-iteration RAW dependencies in the reduction dimension of the kernel. Unroll-and-Jam (Section 3.4) eliminates the cycles wasted due to FPU pipeline stalls, by interleaving the computation of five elements of the result at a time.

We also consider the impact of our optimization passes on registers usage. Each of these passes simplifies either the control flow or memory accesses of the computation, bringing down the number of used registers. However, Fuse

**Table 3.** Our compilation pipeline leverages custom ISA extensions and knowledge of FPU design in order to achieve over 90% FPU occupancy for the MatMul kernel, operating on 1×200 and 200×5 64-bit inputs. Incrementally adding each optimization minimizes and, eventually eliminates, explicit memory operations, while reducing execution time (cycles) and maximizing FPU utilization.

| Optimizations | Allocated Registers (#) | | Assembly Operations (#) | | | | Performance | |
|---|---|---|---|---|---|---|---|---|
| | FP | Integer | Loads | Stores | FMAdd | FRep | Cycles (#) | Occupancy (%) |
| Baseline (for MatMul) | 3/20 | 13/15 | 3 000 | 1 005 | 1 000 | 0 | 40 161 | 2.49 |
| + `Streams` | 3/20 | 11/15 | 1 000 | 1 000 | 1 000 | 0 | 19 165 | 5.25 |
| + `Scalar Replacement` | 3/20 | 10/15 | 5 | 5 | 1 000 | 0 | 4 147 | 24.28 |
| + `FRep` | 3/20 | 9/15 | 5 | 5 | 1 000 | 2 | 4 124 | 24.42 |
| + `Fuse Fill` | 5/20 | 8/15 | 0 | 0 | 1 000 | 1 | 4 130 | 24.5 |
| + `Unroll-and-Jam` | 8/20 | 7/15 | 0 | 0 | 1 000 | 1 | 1 115 | **90.67** |

**Table 4.** Our multi-level compiler backend for Snitch demonstrates the versatility of core concepts in enabling generalized ISA implementation and accelerator development.

| Implementation Features of Our Multi-Level Backend | Concepts |
|---|---|
| Instructions (standard and Snitch) | Operations |
| Instruction operands | SSA values |
| Registers (standard and Snitch SSRs) | Attributes |
| Scoping (based on instruction semantics) | Blocks and Regions |
| Snitch FREP and branch instructions | Control flow dialects |
| Snitch semantics | Custom dialects |
| Target code generation | |
| Register allocation | Progressive lowering |
| Target-specific optimizations | |

Fill increases FP register usage, because two extra variables are required for storing the initial scalar value and performing accumulation. After `Unroll-and-Jam`, register pressure increases to eight FP registers due to the extra interleaved computation. As discussed in Section 3.3, choosing a smaller unroll factor can alleviate increased register pressure.

The combination of high- and low-level passes yields nearly optimal performance when compiling linear algebra micro-kernels using our compiler, reaching 73%-90% FPU utilization across our kernels. We show our ability to leverage the high-level semantics carried by linalg.**generic** to effectively target custom ISA extensions like SSR and FREP.

## 5 Discussion

By modeling our accelerator-specific backend within the MLIR framework, we embraced and materialized its core concepts and principles [53] (Table 4). We encoded the semantics of the standard target ISA and its extensions with operations following the SSA form and used regions to scope the code, avoiding the usual flat representation of assembly languages. In terms of code generation, we applied a progressive lowering approach to target custom accelerator features, adapting high-level, domain-specific information in incremental steps, while adhering to their semantics. Traversing the abstraction levels in this manner has also enabled us to gradually tackle typical compiler backend tasks, such as register allocation. We hope our work will inspire and help shape the path forward towards using structured, domain-specific information in industrial-grade compiler backends for custom accelerators.

## 6 Related Work

We build on top of recent advances in domain-specific compilation, optimization techniques in high-performance computing, and register allocation on IRs in SSA form.

***Compilers.*** Compilers targeting general-purpose CPUs have typically settled in a 3-tier design to decouple implementation concerns and tackle problems independently (e.g., instruction selection and register allocation in the backend), leading to largely monolithic compiler tiers as their sophistication increased. Efforts to enhance composability and modularity—such as separating loop execution schedules from algorithms in polyhedral compilers [18, 37]—have largely failed to permeate mainstream compiler backends. Closer to our work is another micro-kernel compiler [24] based on a DSL [43], which focuses only on matrix multiplication and requires user input to perform scheduling, data transfers and transforms. MoNaCo is an MLIR code generator [68], which emphasizes just-in-time (JIT) compilation speed for the x86-64 ISA, but offers limited support for high-level abstractions, mapping only low-level dialects directly to specific instructions, while avoiding any reliance on the LLVM backend. Our lowering pipeline is automatic and operates on a wide range of linear algebra operations.

A few approaches within the MLIR ecosystem have aimed to generate efficient code with the target micro-architecture in mind, but end up reusing existing general-purpose compiler backends. A case study [20] performs high-level loop transformations in MLIR to match optimizations known to be effective in generating kernels for basic linear algebra subprograms (BLAS) implementations. A similar approach [51] employs an end-to-end C/C++ compilation flow via LLVM IR builtin types. Another technique [72] generates optimized BLAS kernels for ARM-based micro-architectures, focusing on a profile-driven exploration of transformations within the vector dialect. A GPU code generation pipeline [46] ingests high-level dialects and generates LLVM IR of matrix multiplication kernels for NVIDIA GPU tensor cores. To the best of our knowledge, our work is the first MLIR-based compiler backend that captures accelerator-specific abstractions and employs a self-contained code generation scheme.

*Libraries.* Expertly-tuned code for linear algebra and other domains is often shipped in libraries, like oneMKL [9], cuDNN [27] and others [36, 71], typically following a common interface (e.g., BLAS [30]). These libraries offer limited flexibility, using kernel templates, ahead-of-time (AOT) or JIT compilation, to produce tailored library code based on target information (e.g., cache sizes) [5, 41, 71]. LIBXSMM [41] takes this approach to extremes by embedding numerous decisions across various levels of abstraction into its code generation strategy. We reinstate the compiler as the means to flexibly express low-level, target-specific abstractions for generating highly-optimized code.

*DSLs.* A plethora of application-specific languages and accompanying support tooling has been developed over recent years to expose and control aspects of computation (e.g., scheduling, memory placement, etc.) [18, 25, 40, 43, 47, 64, 66, 73, 81], inspired by approaches like Halide [64]. This trend has also led to the adoption of autotuning and analytical frameworks for exploring configuration options [26, 54, 69, 70]. Triton-C [69] extends C with features for tensor manipulation and a builtin autotuner, operating on predefined user-opaque micro-kernels. Inevitably, these DSLs reimplement closely coupled compiler infrastructure [48, 69], rely on general-purpose backends for code generation [43] or require external tools for interoperability [35]. Embedding our approach within MLIR allows us to leverage open, generic and reusable infrastructure, while encapsulating domain and target-specific concepts at the right level of abstraction.

*Decoupled Access-Execute.* Prior work on decoupling data transfers from computation has used best-effort manual [49] and compiler-based transforms [44, 74] in combination with CPU frequency scaling to achieve varying energy and performance objectives. Our work guarantees the separation of access from execution, expressed in the Snitch ISA extensions, as a consequence of lowering high-level operations without requiring a user-specified schedule.

*Register Allocation.* Our approach extends and adapts existing research [21, 39, 62, 76] on SSA-form register allocation by shifting the focus from whole-program basic blocks and $\phi$ functions to structured control flow regions of micro-kernels in support of maximizing hardware utilization.

## 7 Conclusion

The strict separation of frontends and backends in the *hourglass* design of modern general-purpose compilers results in an information bottleneck between high-level code abstractions and targeted novel hardware features. To achieve peak performance, experts often circumvent the compiler with hand-tuned kernels, presenting conceptual *black boxes*, opaque to users and automated analyses. The cost and effort in hand-tuning kernels is exacerbated by a recent proliferation of specialized hardware, driven by manufacturing challenges in modern silicon technologies. In this work, we rethink the compiler backend structure, utilizing a structured, abstraction-driven strategy with multi-level SSA-based IRs. Our MLIR-based prototype showcases a progressive methodology for creating modular and expressive compiler backends that combine domain knowledge with hardware capabilities. We generate assembly for RISC-V targets and Snitch ISA extensions that reaches 90% FPU utilization on representative ML kernels. Our contributions comprise a RISC-V ISA encoding as a multi-level SSA-based IR, a multi-level compiler backend, a tailored register allocator, and a code generation approach for performant linear algebra micro-kernels.

## Acknowledgments

This work has received funding from the European Union's Horizon EUROPE research and innovation program under grant agreement no. 101070375 (CONVOLVE) and was also supported by Research Foundation-Flanders (FWO) under grant 1SE7723N.

## A Artifact Appendix

### A.1 Abstract

This is the supporting artifact for the paper titled "A Multi-Level Compiler Backend for Accelerated Micro-kernels Targeting RISC-V ISA Extensions" as published in CGO 2025. It can be used to reproduce all results presented in the final version of this paper.

### A.2 Artifact Check-List

- **Algorithm:** High-performance micro-kernel compilation for linear algebra operations to the Snitch RISC-V ISA.
- **Program:** MLIR code of linear algebra kernels as represented by the linalg dialect and their corresponding implementation in C source code.
- **Compilation:** Publicly available and included in this artifact: xDSL 0.23.0[1], MLIR 16.0.6 and PULP LLVM [6] 0.12.0 (based on LLVM 12.0.1).
- **Transformations:** mlir-opt tool from the MLIR toolchain to obtain linalg.generic versions for each kernel.
- **Binary:** Bare-metal ELF executables for RISC-V as generated by the workflow in the Docker container image.

---

[1]commit: 902bfbc8

- **Data set:** Randomly generated input sets and precomputed output sets for validation against kernel outputs.
- **Run-time environment:** The Docker container image platform is `linux/amd64`.
- **Metrics:** CPU cycle count, CPU throughput, FPU utilization and register pressure.
- **Output:** Execution logs along with derived plots and tables of Section 4.
- **Experiments:** Preparation and reproduction following instructions in this appendix and using scripts of the artifact.
- **How much disk space required (approximately)?:** All assets require a total of 46 GB of disk space. The experiments directory for the Snitch micro-kernel compiler, input sources, scripts, and logs after executing the full benchmark suite require 43 GB of disk space. The Docker container image containing the RTL simulator, LLVM and MLIR compilation toolchains requires 2.7 GB of disk space.
- **How much time is needed to complete experiments (approximately)?:** Highly dependent on CPU and clocked frequency. The execution for the full benchmark suite requires 1 hour and 45 minutes on an AMD Ryzen 9 5950X 16-core CPU at 4.9 GHz with 62 GiB RAM.
- **Publicly available?:** Yes.
- **Code licenses (if publicly available)?:** Apache License version 2.0 with LLVM Exceptions.
- **Archived (provide DOI)?:** 10.5281/zenodo.14052014.

### A.3 Description

**A.3.1 How Delivered.** The artifact [56] is archived at 10.5281/zenodo.14052014.

**A.3.2 Hardware Dependencies.** A machine with Internet access is required because artifact execution uses `pip` to install Python packages on a virtual environment.

**A.3.3 Software Dependencies.** Working Docker installation. This has been tested on a Linux x86 host machine with Docker 27.3.1.

### A.4 Installation

**A.4.1 Experiments Repository.** Download the experiments repository tarball, navigate to the directory containing the download and extract it:

```
$ tar xvfz riscv-paper-experiments.tar.gz
```

**A.4.2 Docker Container.** Download the Docker image containing the LLVM and MLIR toolchains, navigate to the directory containing the download, load and start it:

```
$ docker load --input snitch-toolchain-artifact.tar.gz
$ docker start snitch-toolchain-artifact
```

### A.5 Experiment Workflow

**A.5.1 Execution of Benchmark Suite.** Run the full benchmark suite by invoking `make` with the `all` target:

```
$ docker run -ti --volume ${PWD}/riscv-paper-experiments:/src \
    snitch-toolchain-artifact:latest bash -c "make -C /src artifact"
```

This command will build the kernels for each experimental flow (our micro-kernel compiler, MLIR and LLVM as described in Section 4.1), execute them with Verilator, process the traces and collate the results in CSV files. The resulting CSV files can be accessed on the host machine (e.g., using `ls`):

```
$ ls -l riscv-paper-experiments/results/
```

Produce the visual elements of Sections 4.2 to 4.4:

```
$ docker run -ti --volume ${PWD}/riscv-paper-experiments:/src \
    snitch-toolchain-artifact:latest \
    bash -c "/src/plots-cgo2025-ae/plot.sh"
```

The resulting files can be accessed on the host machine:

```
$ ls -l riscv-paper-experiments/plots-cgo25-ae/output/
```

### A.6 Evaluation and Expected Result

The experiments are performed on a bare-metal, cycle-accurate simulator for the in-order Snitch CPU, all measurements are deterministic. Hence, all results are expected to closely replicate what is presented in this paper.

### A.7 Experiment Customization

See the `README.md` file, located at the top level of the experiments repository, for instructions on how to run, interpret and process individual experiments.

### A.8 Notes

- Depending on the host machine configuration, docker commands might require elevated privileges (e.g., `sudo`).